\documentclass[aps,prb,twocolumn,floats,showpacs]{revtex4}
\usepackage{graphicx}
\usepackage{amsmath}
\newcommand{\half}{\mbox{\small $\frac{1}{2}$}}
\newcommand{\hide}[1]{}

\begin{document}


\title{Persistent currents in M\"obius strips}

\author{\large Kousuke Yakubo$^1$,
Yshai Avishai$^{1,2,3}$ and Doron Cohen$^{2}$
}

\affiliation{$^1$Department of Applied Physics,
Hokkaido University, Sapporo 060-8628, Japan \\
$^2$Department of Physics and $^3$Ilse Katz Center for Nanotechnology
Ben Gurion University, Beer Sheva, Israel}

\begin{abstract}
Relation between the geometry of a two-dimensional sample and its
equilibrium physical properties is exemplified here for a system
of non-interacting electrons on a M\"obius strip. Dispersion
relation for a clean sample is derived and its persistent current
under moderate disorder is elucidated, using statistical analysis
pertinent to a single sample experiment. The flux periodicity is
found to be distinct from that in a cylindrical sample, and the
essential role of disorder in the ability to experimentally
identify a M\"obius strip is pointed out.
\end{abstract}


\maketitle

\section{Introduction}
\label{sec1}

An important theme in quantum mechanics is to find a relation
between a global geometry of a sample ({\it e.g.}, boundary
conditions) and its physical properties. We address this issue by
comparing flux periodicity of persistent currents in a cylinder
and in a M\"obius strip. The aim is to determine whether the
geometrical (in some sense topological) difference is tangible and
experimentally observable. At zero temperature, the persistent
current $I(\phi)$ in a ring can be expressed as\cite{Imry}
\begin{equation}
I(\phi)=-\frac{\partial E(\phi)} {\partial \phi}=
\sum_{n=1}^{\infty} I_{n} \sin (2 \pi n \phi),
\label{Iphi}
\end{equation}
where $\phi$ is the magnetic flux threading the ring in units of
$\Phi_0=hc/e$, $E(\phi)$ is the ground state energy, and $I_n$ are
the current harmonics.

The current $I(\phi)$ is an anti-symmetric and periodic function
of $\phi$ with period~1. Possible occurrence of smaller flux
periodicity in mesoscopic physics is one of the cornerstones of
weak localization. For the cylinder geometry, conductance
measurements\cite{AASS} and magnetization of $10^7$ copper
rings\cite{Levy} indicate the emergence of periodicity $1/2$. It
is shown to be intimately related to the procedure of {\em
averaging} over disorder realizations and numbers of electrons in
the rings\cite{Gefen,Gilles,AGI,AASS}. Very recently, a
microscopic NbSe$_{3}$ M\"obius strip has been
fabricated\cite{Tanda}. Obviously, in this case, attention should
be focused on a {\it single sample} measurement\cite{Webb} for
which there is no self-averaging.

Let us first mention several intuitive points relevant to the flux
periodicity in the M\"obius strip, based on semi-classical
arguments and geometry\cite{Hayashi}. First, recall that the
periodicity is related to interference between trajectories (such
as Aharonov-Bohm interference between different trajectories or
weak-localization interference between time-reversed paths). In
the cylinder (M\"obius) geometry, an electron moving in the {\em
longitudinal direction} along the ring encircles the system once
({\em twice}) before returning to its initial position. Therefore,
we might expect different flux periodicities of the persistent
current between the two cases. Second, unlike a cylinder which can
be \lq\lq pressed\rq\rq\  into a one-dimensional ring, the
M\"obius strip cannot be pressed into a one-dimensional structure.
This brings in another important factor, namely, the motion of
electrons in the {\em transverse} direction. In a tight-binding
model this motion is controlled by the transverse hopping. If it
is very weak, the twice-encircling property of the M\"obius strip
implies the dominance of even harmonics $I_{2n}$. Contrary, for a
strong transverse hopping, the current in the M\"obius strip is
expected to be effectively similar to that in the cylindrical
strip\cite{Mila}. In the following we are mainly interested in a
regime where the transverse hopping is slightly less than with the
longitudinal one. Third, the role of disorder should be carefully
examined. Weak disorder is not expected to significantly alter
interference between semi-classical trajectories discussed above,
while strong disorder should result in a reduced sensitivity to
the pertinent geometry, due to localization effects. The most
intriguing disorder effect might then be expected in a moderate
strength of disorder which will be used below. The upshot of the
present study is that the periodicity pattern in a M\"obius strip
is remarkably distinct from that of a cylinder, and that disorder
plays a crucial role in making the statistical effect detectable.

\section{Model}
\label{sec2}

A M\"obius strip is modelled by considering a non-interacting
particle in a rectangle of length $L_x$ and width $L_y$, requiring
its wave-function $\psi(x,y)$ to satisfy Dirichlet boundary
conditions in the $y$ direction, and M\"obius boundary conditions
\cite{rmrk} in the $x$ direction:
\begin{align}
&\psi(x,-L_{y}/2) = \psi(x,L_{y}/2) = 0 &&\text{(Dirichlet B.C.)}\ ,\\
&\psi(x{+}L_x,y) = \psi(x,-y) &&\text{(M\"obius B.C.)}\ .
\label{BC}
\end{align}
The quantized wave-numbers are $k_y = (\pi/L_y)n_y$ and $k_x =
(2\pi/L_x)([\half]_{n_{y}}+n_x)$, where $n_y=1,2,\cdots$ and
$n_x=0,\pm1,\pm2,\cdots$. The notation $[\alpha]_{n}$ represents
$\alpha$ for $n=\text{even}$ and $0$ for $n=\text{odd}$. In the
cylinder geometry, Eq.~(3) should be replaced by
$\psi(x+L_{x},y)=\psi(x,y)$, and gives $k_{x}=(2\pi/L_{x})n_{x}$.
Thus, only the $n_y=\text{even}$ eigenstates are affected by the
switch from the conventional cylinder (periodic) boundary
conditions to the M\"obius ones.

In the absence of disorder, the energies of the eigenstates both
in the M\"obius and cylinder strips are given by the formula
\begin{equation}
E_{n_x n_y} =
\epsilon_x\left( k_x-\frac {2\pi\phi} {L_x}\right) + \epsilon_y(k_y),
\label{Enxny}
\end{equation}
where $\epsilon_x$ and $\epsilon_y$ provide the dispersion
relation. Equation (\ref{Enxny}) is rather general for clean
systems. To be more specific, let us model the M\"obius strip by a
tight-binding Hamiltonian. The M\"obius strip is constructed from
a rectangular lattice including $N\times 2M$ sites. The rectangle
is twisted by $180^{\circ}$, and its two sides are connected, such
that longitudinal wire $1$ is attached to wire $2M$, wire $2$ is
attached to wire $2M-1$ and so on. The M\"obius strip so
constructed includes $M$ longitudinal wires with $2N$ sites on
each one. The Hamiltonian is then
\begin{eqnarray}
H_{\text{M\"obius}}=
\sum_{n=1}^{2N} \sum_{m=1}^{M}\Big[
\varepsilon_{nm}c^{\dagger}_{nm} c_{nm}
-t_{1}e^{-2 \pi i \phi/N}c^{\dagger}_{nm} c_{n+1m} \Big]
\nonumber \\
\hspace*{0.5cm}
- t_{2} \sum_{n=1}^{2N} \sum_{m=1}^{M-1}c^{\dagger}_{nm+1} c_{nm}
- \frac{t_{2}}{2} \sum_{n=1}^{2N}c^{\dagger}_{nM} c_{n+NM}
+\text{h.c.}
\hspace*{0.5cm}
\label{HM}
\end{eqnarray}
where $c_{nm}$ is the fermion operator at the site $(n,m)$
($n=1,2,\dots, 2N$, $m=1,2,\dots,M$) and $t_{1}$ and $t_{2}$ are
longitudinal and transverse hopping amplitudes respectively. The
quantity $\varepsilon_{nm}$ is the site energy. Connecting the two
sides of the rectangle without twisting, we obtain a cylindrical
strip which includes $2M$ longitudinal wires composed of $N$
sites. The Hamiltonian of the cylinder is
\begin{eqnarray}
H_{\text{cylinder}}=
\sum_{n=1}^{N} \sum_{m=1}^{2M}\Big[
\varepsilon_{nm}c^{\dagger}_{nm} c_{nm}
-t_{1}e^{-2 \pi i \phi/N}c^{\dagger}_{nm} c_{n+1m} \Big]
\nonumber \\
\hspace*{0.5cm}
- t_{2} \sum_{n=1}^{N} \sum_{m=1}^{2M-1}c^{\dagger}_{nm+1} c_{nm}
+\text{h.c.}
\hspace*{3.0cm}
\label{HC}
\end{eqnarray}
Locally the two Hamiltonians (\ref{HM}) and (\ref{HC}) look the
same. But there is a couple of essential differences between them:
a) The M\"obius Hamiltonian (\ref{HM}) includes an extra term
which describes long range hopping between distant parts of the
$M$th wire \cite{rmrk}. b) While the magnetic phase accumulated
along the longitudinal direction on each link is the same (that
is, $2 \pi \phi /N$), the corresponding number of links is
different ($2N$ for the M\"obius strip and $N$ for the cylinder).

\section{The Spectrum}
\label{sec3}

We first consider a system without disorder, namely,
$\varepsilon_{nm}=0$. The dispersion relation for an electron in
the M\"obius strip reads,
\begin{eqnarray}
E_{n_x n_y} =
&-& 2t_1 \cos \Bigg[\frac{2\pi}{N}\left( \Big[ \frac{1}{2}\Big]_{n_{y}}
+n_x-\phi\right)\Bigg] \nonumber \\
&-& 2t_2 \cos \left(\frac{\pi}{2M{+}1} n_y \right)\ ,
\label{Enxny1}
\end{eqnarray}
where $n_x=1,\cdots,N$ and $n_y=1,\cdots,2M$. Defining new indexes
$k=[1]_{n_{y}}+2n_{x}$ and $q=[\half]_{k}+n_{y}/2$, one obtains a
more suggestive form,
\begin{eqnarray}
E_{kq} =
&-&2t_1 \cos \Big[\frac{\pi}{N}(k-2\phi)\Big] \nonumber \\
&-&2t_2 \cos \Big[\frac{\pi}{2M+1} \left(2q-[1]_{k}\right) \Big],
\label{EkqM}
\end{eqnarray}
where $k=1,\cdots,2N$ and $q=1,\cdots,M$. It is instructive to
compare it with the energy in the cylinder geometry, $E_{kq}=
-2t_1 \cos \big[\frac{2 \pi}{N}(k-\phi)\big] -2t_2 \cos
\left(\frac{\pi}{2M+1} q \right)$, where $k=1,\cdots,N$ and
$q=1,\cdots,2M$. Despite the apparent similarity between these two
spectra, there are at least two important differences. First, the
combination of flux and longitudinal momentum is distinct, namely,
it is $k-\phi$ for the cylinder and $k-2\phi$ for the M\"obius
strip. For a small ratio $t_2/t_1$ this might affect the
periodicity of the current\cite{Mila}. Second, the mini-band
structure is different.
\begin{figure}[tttt]
\includegraphics[width=\hsize]{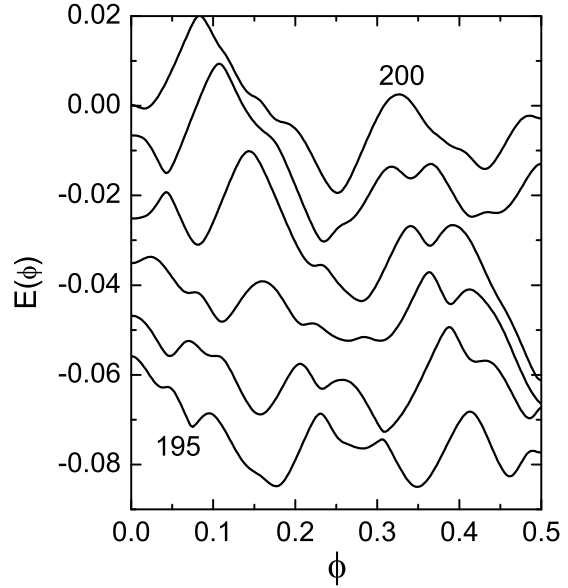}
\caption{Single-particle energy spectrum as a function of flux
threading the M\"obius ring. The $195$th-$200$th energy levels are
shown. The parameters are $N=20$, $M=10$, $t_{2}=0.5$, and
$W=0.5$. Energies are measured in units of $t_{1}$.}
\label{fig1}
\end{figure}

We now turn to elucidate the current in disordered M\"obius
strips. The random numbers $\varepsilon_{nm}$ are assumed to be
uniformly distributed over the range $-W/2\le \varepsilon_{nm}\le
W/2$, where $W$ represents the strength of disorder. The
Hamiltonian Eq.~(\ref{HM}) [or Eq.~(\ref{HC})] is treated
numerically. As an example, the evolution of single-particle
energies with flux in a disordered M\"obius strip with $N=20$ and
$M=10$ is shown in Fig.~1. The parameters are $t_{2}/t_{1}=0.5$
and $W/t_{1}=0.5$. The pattern of avoided crossing turns out to be
remarkably different from that for a cylinder (see ref.~5 figure 1
therein). It must then be reflected in the behavior of persistent
currents.

\begin{figure}
\includegraphics[width=8cm]{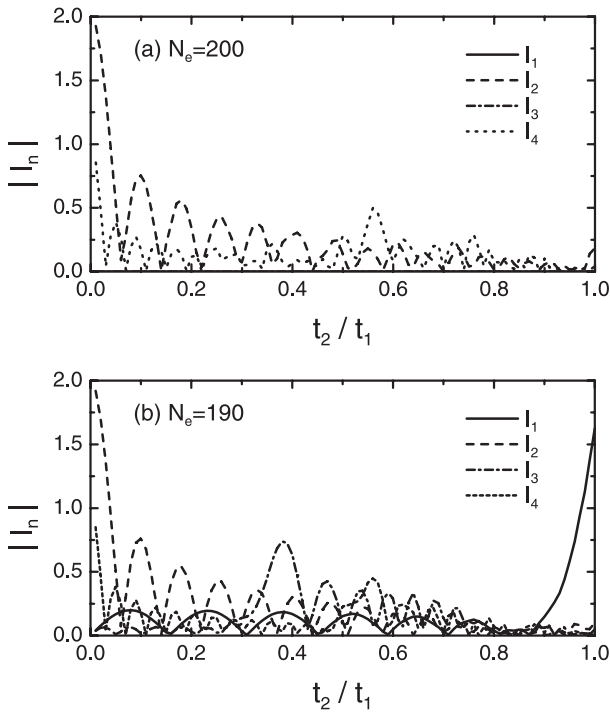}
\caption{Fourier components of the persistent current for the
clean M\"obius strip as a function of the transverse-hopping
energy (a) at the half-filling and (b) below the half-filling. The
size of the M\"obius strip is given by $N=20$ and $M=10$.}
\label{fig2}
\end{figure}
The first stage of the analysis is an inspection of the typical
values of $I_n$, aiming in determination of their dependence on
the ratio $t_2/t_1$. As expected, in the absence of averaging we
find typical $I_1$ dominance in case of the cylinder geometry
irrespective of the $t_2/t_1$ ratio. For the M\"obius geometry the
emerging picture is quite different. Figure 2 shows the Fourier
components of the persistent current for a clean M\"obius strip as
a function of the ratio $t_{2}/t_{1}$ at and below the
half-filling ($N_{e}=200$). For small ratios ($t_2/t_1 < 0.1$) we
find, as can be naively expected, $I_2$ dominance.
The expected effect of averaging in the cylinder case is to
emphasize the $I_2$ contribution, while in the M\"obius case
the expected effect is to emphasize the $I_4$ contribution.
For clean M\"obius strip the $I_{n}$ with odd $n$,
as a function of the number of electrons $N_{e}$,
is anti-symmetric around half-filling.
Therefore $I_{1}$ and $I_{3}$ completely vanish [Fig.~2(a)].
See further discussionin Sect.~V.
To avoid this particularity at the half-filling,
we display in Fig.~2(b) also the case where
the number of electrons ($N_{e}=190$) is below
half-filling.
For large ratios ($t_2/t_1 > 0.8$) we observe in Fig.~2(b)
a cylinder-like regime where there is typically $I_1$ dominance.
This is because the strong transverse hopping changes the periodicity
of the M\"obius strip to that of the conventional cylinder.
The somewhat unexpected observation is that there is a distinct wide
intermediate regime ($0.1 < t_2/t_1 < 0.8$) where $I_1$, $I_2$,
$I_3$ and $I_4$ are all comparable. This is the regime which is of
experimental relevance.
The expected effect of averaging in this regime is to emphasize
both the $I_2$ and the $I_4$ contributions.

\section{Statistical Analysis}
\label{sec4}

The problem arising in the analysis of persistent currents in
disordered M\"obius strips is how to characterize the statistics
of the calculated data. It was already pointed out that essential
properties of observables result from the averaging procedure and
the nature of the underlying statistical ensemble
\cite{Gefen,Gilles,AGI}. On the other hand, fabrication of a
M\"obius strip requires an outstanding effort\cite{Tanda}, and
hence, anticipated measurements of the persistent current would
probably be performed on a single sample. Thus, somewhat
unfortunately, the important results reported therein and the
powerful calculation methods based on super-symmetry might be less
useful for {\em single-sample} experiments since there is no
averaging.

What is then the most efficient way to present our calculated
results? The answer is provided by elementary statistics. An
experimental result consists of a set of $K$ measurements
$I(\phi_{i})$, $i=1,2,\dots,K$ performed on a given sample. This
sample is taken out of an ensemble of M\"obius strips with
different disorder realizations, electron numbers $N_{e}$, aspect
ratios, {\em etc}. The set $\{ I(\phi_{i}) | i=1,\dots,K  \}$ can be
regarded as an instance of a random vector in a $K$ dimensional
space. Alternatively, this instance can be represented by the
current harmonics $(I_{1},I_{2},\dots)$ defined via
Eq.~(\ref{Iphi}). For our purpose it seems adequate to keep only
the first 4~harmonics. The relevant statistical ensemble is then a
set of \lq\lq points\rq\rq\ $(I_{1},I_{2},I_{3},I_{4})$ in
four-dimensional probability space, each point corresponds to a
possible experimental measurement of the current on the {\it
entire} $\phi$ interval. Let us denote the number of points within
an infinitesimal four-dimensional volume element by
$P(I_{1},I_{2},I_{3},I_{4})dI_{1}dI_{2}dI_{3}dI_{4}$. The
distribution function $P$ is normalized to ${\cal N}$, the total
number of members in the ensemble. The most probable (typical)
experimental result is then determined by the quadruple
$I_{1},I_{2},I_{3},I_{4}$ at which $P$ is maximal. Another
quantity, which seems more informative and easy to analyze, is the
distribution
\begin{equation}
p_{n}(I_{n})=\int_{0}^{\frac {1} {2} |I_{n}|}
P(I_{1},I_{2},I_{3},I_{4}) \prod_{m \ne n} d|I_{m}|.
\label{pn}
\end{equation}
This corresponds to the possibility of finding a sample whose
current $I(\phi)$ is approximately described by $I(\phi) \approx
I_n \sin(2\pi n\phi)$. (For a sample counted by $p_{n}(I_{n})$,
all the harmonics other than $I_{n}$ are at most half of $I_{n}$
in magnitude). The number of members in the ensemble that exhibit
$I_n$ dominance is therefore ${\cal N}_{n}=\int_{0}^{\infty}
p_{n}(I_{n})d|I_{n}|$. If ${\cal N}_{n} > {\cal N}_{m}$ for any $m
\ne n$, the typical periodicity of $I(\phi)$ is dominantly $1/n$.
\begin{figure}[tttt]
\includegraphics[width=\hsize]{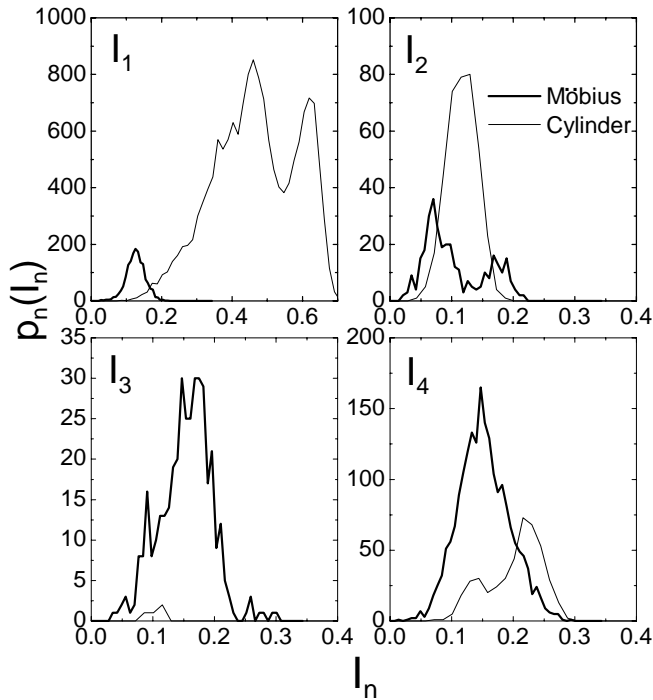}
\caption{The distributions $p_{n}(I_{n})$ ($n=1,2,3,4$) defined by
Eq.~(\ref{pn}) for the cylinder and M\"obius ensembles. The
numbers of members with $I_n$ dominance are ${\cal N}_{1}=15,829$,
${\cal N}_{2}=382$, ${\cal N}_{3}=4$, and ${\cal N}_{4}=439$ for
the cylinder ensemble, and ${\cal N}_{1}=1,562$, ${\cal
N}_{2}=336$, ${\cal N}_{3}=384$, and ${\cal N}_{4}=1,992$ for the
M\"obius ensemble.}
\label{fig3}
\end{figure}
In actual calculations, we assume that the lattice structure, the
aspect ratio, and the strength of disorder are fixed, and that the
temperature is very low. Then, two quantities are still
fluctuating, namely, the filling factor (or the electron number
$N_{e}$) and the specific realization of disorder. We generate an
ensemble of ${\cal N}={\cal N}^{a}{\cal N}^{b}$ members
corresponding to ${\cal N}^{a}$ consecutive values of $N_{e}$,
usually around half filling, and ${\cal N}^{b}$ realizations of
disorder for each one of them.  Actually, for our systems of size
$N=20$, $M=10$ with $t_{1}=1$, $t_{2}=0.5$, and $W=0.5$, we take
$150 \le N_{e} \le 250$, hence ${\cal N}^{a}=101$ and ${\cal
N}^{b}=250$, so that ${\cal N}=25250$. The distributions
$p_{n}(I_{n})$ for the cylinder and M\"obius ensembles are shown
in Fig.~3.

\section{Main Observations}
\label{sec5}

The most striking result that can be deduced from Fig.~3 is the
essential reduction of ${\cal N}_{1}$ for the M\"obius ensemble
compared with the cylinder one. For the present ratio $N/2M=1$,
there is also a strong tendency towards $\Phi_{0}/4$ periodicity,
since ${\cal N}_{4} > {\cal N}_{m \ne 4}$ for the M\"obius
ensemble. This result is intriguing, because here we have no
averaging procedure which is crucial to get the $1/2$ periodicity
in cylindrical strips. However, this $1/4$ periodicity emerges
only for the specific ratio $N/2M=1$. We have calculated the
distributions $p_n(I_n)$ for M\"obius strips with several aspect
ratios. The value of ${\cal N}_{n}$ depends on the aspect ratio.
No specific $n$ gives prominent ${\cal N}_{n}$ independently of
the aspect ratio. On the other hand, the collapse of $I_{1}$
dominance in the M\"obius ensemble is robust and persists in
systems with different ratios $N/2M$ as well. We can safely say
that ${\cal N}_{1}$, ${\cal N}_{2}$, ${\cal N}_{3}$, and ${\cal
N}_{4}$ become all comparable in the M\"obius ensemble.

\begin{figure}
\includegraphics[width=8.5cm]{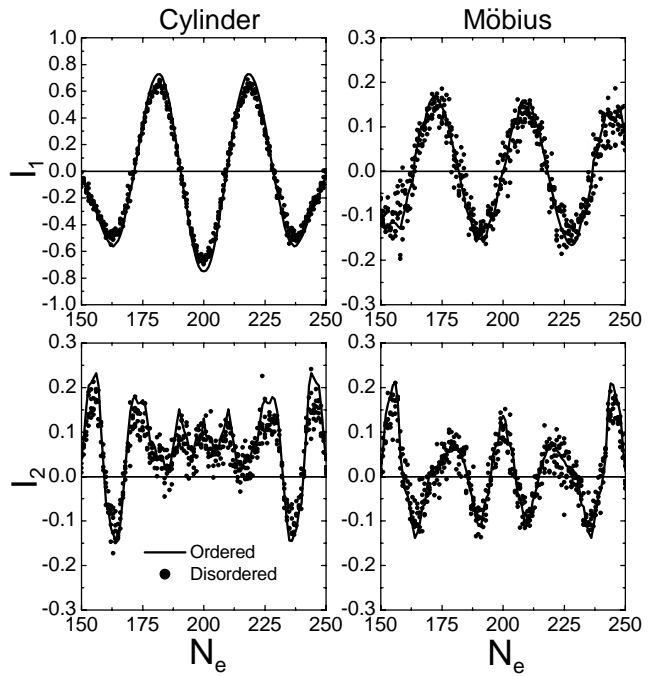}
\caption{$I_{1}$ and $I_{2}$ as a function of $N_{e}$ for the
ordered (solid line) and disordered (dots) systems. Parameters
describing the systems are the same with those for Fig.~3.}
\label{fig4}
\end{figure}
The natural question that comes to mind is whether this result is
a consequence of the M\"obius geometry or, rather, it is due to
the presence of disorder. In order to answer this question, we
have performed the calculation of $P^0(I_1,I_2,I_3,I_4)$ for a
\lq\lq clean\rq\rq\ M\"obius ensemble (without disorder, only
$N_e$ is being changed). We found out that the probability to find
any $I_n$ dominance is extremely small. The immediate conclusion
is that disorder is essential for the identification of M\"obius
strips via $I_{n>1}$ dominance. Does this mean that interference
or weak-localization effects due to the presence of disorder is
important? To clarify this point, we should understand how
$P^0(I_1,I_2,I_3,I_4)$ is modified by disorder. The distribution
$P^0(I_1,I_2,I_3,I_4)$ is, in fact, a function defined on a one
dimensional curve $[I_1(N_e),I_2(N_e),I_3(N_e),I_4(N_e)]$ in
$(I_1,I_2,I_3,I_4)$ space. For this reason, it is unlikely to find
a sample where one of the $I_{n}$ is dominant. The effect of
disorder is to give some \lq\lq thickness\rq\rq\ to this curve
(see Fig.~4). Taking into account that the amplitudes of
$I_n(N_e)$ for M\"obius strips are all comparable, the thickness
gives a finite probability to find samples where one of the $I_n$
is dominant. On the contrary, in the case of cylindrical strips,
the amplitude of $I_{1}(N_{e})$ is overwhelmingly larger than
those of $I_{n\ne 1}(N_{e})$, which makes it unlikely to find
$I_{n\ne 1}(N_{e})$ dominated samples even if we take the
statistical effect of disorder into account. We should note here
that the function $I_{n}$ with odd $n$ for the clean M\"obius
strip is an even function around the half-filling ($N_{e}=200$)
and an odd function for odd $n$, while the function $I_{n}$ for
arbitrary $n$ is an even function in the cylinder case.

Our findings regarding ${\cal N}_{n}$ for the M\"obius ensemble
are based on the fact that the amplitudes of $I_{n}(N_{e})$
are all comparable for M\"obius strips. As we have observed
in Fig.~2, this is a robust statistical property in the
intermediate regime $0.1<t_{2}/t_{1}<0.8$.
The choice $t_{2}/t_{1}=0.5$ above, provides typical
results for $p_{n}(I_{n})$ and ${\cal N}_{n}$ in case
that $t_{2}/t_{1}$ is within this distinct regime.

\section{Conclusions}
\label{sec6}

We have studied the persistent currents of non-interacting
electrons in M\"obius strips. The spectral properties for a clean
system were found analytically, and the effect of disorder on the
currents was analyzed numerically. We have found that disorder is
quite essential for the identification of M\"obius strips. The
issue of disorder averaging is not relevant for single sample
experiments, and hence, special care is required for statistical
analysis of the current harmonics. The fingerprint of the M\"obius
geometry is an enhanced probability to find samples in which
$I_n$, with $n>1$ dominates. This should be contrasted with the
case of cylinder geometry, where there is a clear $I_{1}$
dominance. The above assertion regarding the fingerprint of the
M\"obius geometry is correct provided the effect of disorder is
properly taken into account.

We would like to thank T. Nakayama for very helpful discussions.
One of the authors (Y.A.) was supported by the Invitation
Fellowship for Research in Japan (Short Term) of the Japan Society
for the Promotion of Science. Numerical calculations in this work
have been mainly performed on the facilities of the Supercomputer
Center, Institute for Solid State Physics, University of Tokyo.


\end{document}